# Analysis of the Educative Reform in the Secondary School Education in Mexico and its implications in Science II in the new curriculum

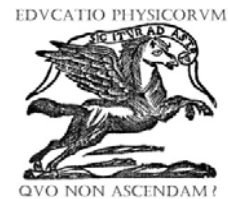

# Análisis de la Reforma Educativa en la Educación Secundaria en México e implicaciones del nuevo plan de estudios en la materia de Ciencias II

**Alfonso Cuervo**[1*], **César Mora**[1] **y R. García-Salcedo**[1]
[1]*Centro de Investigación en Ciencia Aplicada y Tecnología Avanzada.*
*Instituto Politécnico Nacional. Av. Legaria 694, Col. Irrigación, CP 11500, México D. F.*

[*]*Instituto de Humanidades y Ciencias.*
*Puente de Piedra 29A. Col. Toriello Guerra, Tlalpan 01450, México D.F.*

**E-mail:** acuervo@cienciamia.com, cmoral@ipn.mx, rigarcias@ipn.mx,

**Resumen**
En México, la Reforma Educativa que ha ido entrando en vigor gradualmente en la Educación Secundaria desde el año 2005, ha eliminado algunas materias y creado otras que concentran en un solo grado de estudios los temas de la educación básica, despertando una serie de cambios que representan un reto para los docentes, quienes además de adaptarse a nuevas cargas académicas enfrentan un enfoque diferente en el tratamiento de los temas. Más allá de los conflictos que toda reforma pueda generar, el presente artículo sin constituirse en un análisis político o social, aborda algunas de las diferencias entre autoridades de la Secretaría de Educación Pública (SEP) y el magisterio, concentrándose en las implicaciones del nuevo plan de estudios de la materia de Ciencias II (Física) y plantea la necesidad de diseñar una serie de secuencias didácticas para lograr la enseñanza de algunos conceptos físicos en la escuela secundaria.

**Palabras clave:** Educación, educación secundaria, currículo y evaluación.

**Abstract**
In Mexico, the educative reform that has been taking effect gradually in the secondary school education from 2005, has eliminated some subjects and created other that they concentrate in a single degree of studies the themes of the basic education, waking up changes that represent a challenge for teachers, who besides are adapted to new academic loads, they face an approach different in the treatment from the new themes. Beyond the conflicts that all reform can generate, the present paper is not a political or social analysis, but it approaches some differences between authorities of Secretaría de Educación Pública (Public Education Secretary) and teachers, studying the implications of the new curriculum of Sciences II (Physics). Additionally, it is considers the design of didactic sequences to teach some physical concepts in the secondary school by active learning.

## I. INTRODUCCIÓN

En México, en 1993, se declaró como obligatoria a la escuela secundaria como última parte de la denominada educación básica. Para conseguir este carácter de obligatorio, el Estado se compromete a proporcionar las condiciones para que cualquier alumno egresado de la escuela primaria pueda acceder y permanecer en la escuela secundaria hasta finalizarla y ofrecer a los alumnos oportunidades formales para adquirir y desarrollar conocimientos, habilidades, valores y competencias básicas que se requieren para seguir hacia una educación superior o bien para incorporarse al mercado de trabajo.

La población de estudiantes de la Educación Secundaria está entre los 12 y los 15 años, una etapa de la vida en la que ocurren cambios muy importantes, fisiológicos, cognitivos, emocionales y sociales. Así, para tener un gran impacto formativo en la vida de los estudiantes es necesario que las personas que están interesadas en su educación se ocupen en comprender y caracterizar al adolescente que recibe, y de definir con precisión lo que la escuela le ofrece a sus estudiantes, para quienes las transformaciones y la necesidad de aprender nuevas cosas son una constante.

Por lo anterior, la SEP consideró necesario llevar a cabo una Reforma Educativa para que la participación de los alumnos dentro del salón de clases fuera más activa, por lo que promueve la convivencia y aprendizajes en ambientes más colaborativos y desafiantes; posibilita una transformación de la relación entre maestros y alumnos, y facilita la integración de los conocimientos que los estudiantes adquieren en las distintas asignaturas y este proyecto de investigación toma en cuenta estos aspectos.

Los documentos oficiales señalan que el estudio de las ciencias en la educación secundaria debe estar encaminado a que los estudiantes consoliden una formación científica básica que les permita comprender; reflexionar; tener curiosidad, crítica y escepticismo; investigar; opinar; decidir y actuar. De la misma forma, que reconozcan que el conocimiento científico siempre está en constante cambio, el cual es producto de muchas mujeres y hombres de diferentes culturas.

El Plan de Estudios 2006 de la educación secundaria [1], tiene entre sus orientaciones didácticas para el mejor aprovechamiento de los nuevos programas de estudio los siguientes:
1. Incorporar los conocimientos previos de los alumnos.
2. Promover el trabajo grupal y construcción colectiva del conocimiento.
3. Optimizar el uso del tiempo y del espacio.
4. Seleccionar materiales adecuados.
5. Impulsar la autonomía de los estudiantes.
6. Evaluación.

Como parte fundamental del programa de educación nacional 2001- 2006 en noviembre de 2002, se dio a conocer el primer borrador del documento base de la Reforma Integral de la Educación Secundaria (RIES) a través de la Subsecretaría de Educación Básica y Normal (SEByN)[1].

Este primer documento expuso datos en los que se valoró la eficacia de la educación secundaria, considerando las oportunidades para la permanencia de los estudiantes en las escuelas, la deserción, el bajo aprovechamiento académico, los resultados en el Programa Internacional para la Evaluación del Estudiante (PISA), las condiciones históricas, institucionales y escolares asociadas a estos resultados y, por último, abordó los propósitos, características y premisas que, desde la perspectiva de la subsecretaría habrían de orientar el proceso de reforma.

Así comenzaría una serie de diálogos y debates entre todos los actores involucrados (autoridades de la SEP, maestros y su sindicato), con la finalidad de exponer sus posturas y tratar el por qué, para qué y cómo de la transformación de la educación secundaria. Entre los primeros cambios que tendría este documento sería el de quitar la palabra Integral conociéndose sólo como Reforma de la Educación Secundaria (RES).

El desarrollo del Plan de Estudios y documentos relacionados con esta RES, contó con la participación de maestros y directivos de las escuelas secundarias de todo el país, colaboraron también en su creación especialistas en los contenidos de las diversas asignaturas que conforman el plan de estudios[2].

La RES comenzó a implementarse en algunas escuelas en el año 2005 como una primera etapa de prueba y entró en vigor en el primer grado en todas las escuelas secundarias en 2006.

Para el ciclo escolar 2007-2008 se incorporó el segundo grado quedando únicamente el tercer grado de secundaria con el plan de estudios anterior y actualmente, en el ciclo escolar 2008 – 2009 que dio comienzo el día 18 de agosto de 2008 está presente en la totalidad de la Educación Secundaria.

## II. REVISIÓN CRÍTICA DEL PLAN DE ESTUDIOS DESDE LA ACADÉMIA

Uno de los primeros conflictos entre docentes y autoridades fue justamente el hecho de que a decir de los maestros inconformes, no fueron consultados en su totalidad [3][3].

Desde sus inicios, la RES ha sido aplaudida por unos y criticada por otros, especialmente por aquellos docentes que perdieron horas de clase, pues la RES contempla una distribución de la carga académica diferente que ha dejado fuera, en el primer grado de Secundaria a las materias de Historia, Formación Cívica y Ética, e Introducción a la Física y a la Química.

Por otro lado, existe la creación de nuevas materias como son: Orientación y Tutoría[4] y las asignaturas estatales que persiguen el fortalecimiento de la identidad regional y el aprecio de la diversidad del país.

---

[1] Cita de referencia [2]: *"la Subsecretaría de Educación Básica y Normal (SEByN) ha elegido adoptar una estrategia para el diseño y planificación del cambio que, en principio, asegure que todos los actores involucrados en el proceso de reforma de la educación secundaria tengan una comprensión común de sus propósitos y se vean a sí mismos trabajando para su consecución."*

[2] Proceso de Construcción. Reformulación de las propuestas.
La difusión de las versiones elaboradas hasta junio del 2004 dio lugar a una serie de intercambios con maestros, investigadores y otros interlocutores que con sus comentarios, críticas y aportaciones contribuyeron a la elaboración de versiones renovadas, las cuales fueron concluidas entre febrero y mayo del 2005. Mención especial requiere el trabajo realizado con dos instancias:
El Comité Interinstitucional de Historia, que participó con la Secretaría de Educación Pública en la elaboración de los programas de Historia.
http://www.ries.dgme.sep.gob.mx/doc/procesos/pchistoria.pdf .
La Academia Mexicana de Ciencias, que contribuyó de manera destacada en el conjunto de asignaturas del currículo.
http://www.reformasecundaria.sep.gob.mx/doc/proc/proconstru.pdf .

[3] *"…Tanto la opiniones de los docentes como el documento elaborado por la Academia Mexicana de Ciencias fueron tomados en cuenta de manera muy parcial y limitada ignorando los comentarios de fondo…"*.
En: Comentarios a los programas de Ciencias I, II y III, sección: Problemas de la puesta en marcha, p. 1460.

[4] Cita de la referencia [4]: *"La tutoría es un espacio curricular de acompañamiento, gestión y orientación grupal, coordinado por una maestra o un maestro, quien contribuye al desarrollo social, afectivo, cognitivo y académico de los alumnos, así como a su formación integral y a la elaboración de un proyecto de vida."*

Para el Distrito Federal, cuya identidad regional resulta sumamente difícil de definir dado que todos los ciudadanos venimos o tenemos familiares de otra región del país, la materia estatal corresponde a la llamada Aprender a Aprender que brinda estrategias que fortalecen las habilidades para el estudio y el aprendizaje en los alumnos de primer grado.

Cabe aquí señalar que la asignatura estatal del plan de estudios 2006 para la educación secundaria tiene su antecedente en la asignatura opcional del plan de estudios 1993 del mismo nivel.

Siguiendo el acuerdo secretarial 384 de la Ley de Educación, las autoridades educativas locales tienen la facultad de proponer contenidos regionales para la consideración y, en su caso, autorización de la Secretaría de Educación Pública (SEP).

La materia de Expresión y Apreciación Artística se ha transformado en la materia de Artes. Las de Biología y Educación Ambiental se han concentrado en un solo curso titulado Ciencias I, estudiado en el primer grado. Los cursos de Física y Química que eran estudiados tanto en segundo como en tercer grado de Secundaria, ahora son transformados en las materias de Ciencias II (Física) y Ciencias III (Química).

De esta manera, en lo referente a Ciencias, en primer grado se estudia un enfoque en Biología, en segundo grado el enfoque es en Física y en tercer grado se dará el estudio de la Química. Esta situación, desde luego ha desatado la preocupación de los docentes, pues independientemente de que sean más las horas de estudio a la semana en comparación con el plan de estudios anterior, (establecido en 1993), cuentan con un solo ciclo escolar para impulsar en sus alumnos el interés por su materia y aunque los planes de estudio contemplan una interrelación, la incertidumbre y la desconfianza de los docentes son predominantes en la actualidad.

Cabe señalar que en las secundarias mexicanas, existen tanto escuelas de gobierno como particulares. Para el caso de las primeras los docentes que llegan a trabajar en ellas, son personas preparadas en las escuelas superiores o normales, por lo que cuentan con una especialización en la materia que impartirán. En el ramo de las escuelas particulares, se contrata a profesionistas que tienen un perfil de preparación afín con la materia que impartirán. Por lo que un ingeniero en electrónica, por ejemplo, puede impartir las materias de Matemáticas, Informática y Ciencias II (Física). Esto significa en otras palabras, que no está garantizada la especialización de la planta docente.

## III. DIFUSIÓN, CONTROL Y SEGUIMIENTO

La SEP ha elaborado una campaña fuerte para difundir su proyecto de reforma a través de las tradicionales juntas de Academia; los inspectores de zona, los jefes de clase y los coordinadores respectivos han venido convocando a profesores provenientes tanto de escuelas públicas como privadas, a trabajar en una o dos sesiones en la modalidad de taller. Divididos por áreas, los docentes han recibido información de las autoridades de la SEP sobre los nuevos planes de estudio y las metodologías de trabajo que ahora se deben implementar.

A través de circulares y la vía telefónica, se da aviso a los directores de escuela sobre las fechas y horarios en los que la SEP convoca a las juntas de academia, se seleccionan diferentes planteles distribuidos por zonas escolares para que funcionen como sedes de las diferentes academias. Los docentes reciben el aviso de junta de parte de sus directores y ubican la escuela donde deberán presentarse.

Desde luego, este movimiento masivo de docentes fuera de sus centros de trabajo y la ocupación de las escuelas para albergar a los convocados a junta de academia, provoca la suspensión de las labores académicas, rompe los planes y horarios de clase y provoca que no se cumplan los 200 días de clase fijados por la misma SEP.

Cabe señalar que la SEP, dentro de su campaña de información ha elaborado una serie de materiales impresos, antologías y guías de trabajo, con la finalidad de distribuirlos de manera gratuita a los docentes durante las juntas académicas, pero el descontrol administrativo y la premura con que se organizan estos eventos, provoca que no todos los docentes se enteren de la sede que les corresponde, o se presenten más docentes de los esperados en una sede, haciendo que los materiales sean escasos.

Las situaciones arriba descritas provocan apatía, desinterés y hasta frustración en docentes que no alcanzan materiales impresos y que sienten que al presentarse a las juntas de academia, pierden un valioso día de trabajo frente a sus grupos cotidianamente rezagados.

Estos materiales, a los que se hace referencia, han sido creados por grupos de autores, revisores y lectores para dar a conocer la reforma a los docentes, a través de talleres dirigidos por inspectores y jefes de enseñanza quienes a su vez, primeramente han tenido reuniones de capacitación con sus coordinadores estatales[5].

La reforma ha llegado al aula mediante un proceso vertical de transmisión de información de autoridades a maestros por la vía de la "capacitación".

Como medidas de control y de demostración de que el docente realiza una planeación de sus actividades, la SEP ha venido solicitando a los docentes año con año una serie de documentos cuyo formato puede tener variaciones entre zonas escolares y escuelas, pero que básicamente son los mismos y con el paso de los años desde la anterior reforma de 1993 los docentes identifican plenamente.

---

[5] En el Boletín informativo de la Dirección de Ciencias Naturales, DGDC/SEP. Número 5, Noviembre, 2007 (disponible en la página www.reforma.sep.gob.mx ) se dice que el objetivo de estas reuniones ha sido: *Brindar herramientas al personal responsable de la capacitación y asesoría de los docentes que atienden 2° grado, a fin de que profundicen en el conocimiento de los programas de estudio, planteen y resuelvan dudas surgidas durante las acciones de capacitación y proporcionar estrategias para fortalecer su función de asesoría; asimismo, aportar elementos a los supervisores responsables del seguimiento a las escuelas en el marco de la Generalización de la Reforma de Educación Secundaria.*

Estos documentos fueron conocidos como el *Plan de Trabajo Anual*, (mostrado en la tabla I), formato donde el docente planteaba los propósitos, estrategias, recursos e instrumentos de evaluación para cada tema y el *Avance Programático*, documento tipo cronograma que de manera más detallada señalaba el tema por abordar en cada sesión de clase.

**TABLA I.** Campos principales del Formato para Plan de Trabajo Anual hasta antes de la RES.

| Escuela: | Profesor: | Asignatura: | Grado: |
|---|---|---|---|
| Propósitos | Estrategias | Recursos | Procedimiento e Instrumentos de evaluación |
|  |  |  |  |
| Firmas Dirección del Plantel y Docente responsable |  | Vo Bo SEP | Sellos |

Hoy, la Reforma Educativa ha traído consigo nuevos formatos. Los docentes ya acostumbrados a trabajar con documentos de cierto tipo se enfrentan a nuevos formatos. Pese a que la SEP ha instruido en el llenado de éstos en los talleres de actualización, para los docentes más ajenos a la documentación y el papeleo oficial, cambiar el concepto de propósitos por el de aprendizaje esperado o generar secuencias de clases se convierte en una inconformidad más contra la reforma.

Algunos docentes han comentado, en las reuniones de trabajo, que la pobre supervisión que hacen las autoridades de la SEP en estos formatos ha desencadenado el que cada año se repita prácticamente la misma planeación, haciendo cambios de fechas únicamente. El papeleo para estos docentes pasa a ocupar un lugar secundario dentro de la impartición de sus respectivas clases.

El llamado a la conciencia de los docentes sobre la importancia de una buena planeación de clases es aún estéril en muchos casos

Algunos comentarios recabados, reflejo de lo anterior, se muestran a continuación:

*"Cada ciclo escolar es lo mismo al principio de año, llenar el PTA (Plan Anual de Trabajo), entregarlo a la dirección y no volverlo a ver, hasta el siguiente ciclo"*

Docente 1

*"Entregué mi planeación y sé que la dirección la hizo llegar a la SEP, pero los comentarios hechos por el jefe de clase a mi planeación nunca me fueron entregados"*

Docente 2

*"No tiene mucho caso planear si mi hora de clase es la que usan para todas las actividades"*

Docente 3

Este último comentario hecho por un docente, es en relación a que en las escuelas se organizan durante las horas de clase, una serie de actividades extracurriculares como preparación de festejos, ceremonias, muestras de ciencia, ensayos de diferentes eventos, etc.

Con la finalidad de hacer de la Reforma Educativa una acción cotidiana y no una simple verbalización de buenas intenciones, las autoridades de la SEP han advertido en las juntas de academia que comenzarán con más ahínco a supervisar el que dichas planeaciones sean congruentes con los programas, verificando que exista una estrecha relación entre los aprendizajes esperados propuestos en los planes de estudio de la SEP y las estrategias de clase propuestas por los docentes.

Dicha supervisión va más allá de la revisión de documentos y contempla el seguimiento de clases en el aula por parte de jefes de clase e inspectores.

La presencia de las autoridades de SEP en las escuelas no es nueva, pues ya se venía haciendo antes de la RES, pero tiene un enfoque diferente al observar la implementación de las nuevas políticas institucionales.

Los jefes de clase e inspectores se presentan en las escuelas sin aviso previo. Los docentes desconocen el día y la hora en que serán supervisados frente a sus alumnos. Los jóvenes estudiantes en ocasiones no saben si la autoridad de la SEP está evaluando a su profesor o a ellos mismos o ambos.

En algunas sesiones de supervisión, es sabido que el docente cambia su metodología y trato hacia los alumnos cosa que ha llegado a ser evidenciada por los mismos alumnos ante las autoridades. Todo esto trae consigo alteraciones en la vida escolar tanto de docentes como alumnos.

Un par de comentarios extraídos del quinto informe nacional[6] citados a continuación, ilustran la situación arriba señalada.

*En un par de ocasiones en que ingresé al aula donde se impartía la clase, los alumnos hicieron escarnio del profesor en voz alta, preguntándole: —¿También hoy haremos un resumen?, porque usted es lo único que sabe ponernos para trabajar.*
*El profesor alza el tono de la voz, pasa lista y pega una lámina de papel bond en el despintado pizarrón. Los alumnos le increpan: —¿Por qué hoy no nos dicta como siempre hace?... seguramente porque tiene visita de los de la Reforma.*

Registro de observación de un supervisor

*Todo lo que hicimos en las clases de Ciencias II fue simulación, el profesor nos había dicho la forma en que debíamos de comportarnos y de trabajar; además pidió que al alumno que entrevistaran dijera que siempre trabajaban así.*

Comentario de un Alumno

---

[6] Quinto Informe Nacional. Seguimiento a las Escuelas. SEP 2007. Disponible en formato PDF en
http://www.reformasecundaria.sep.gob.mx/pdf/seguimeintopei/quinto_informe_nacional.pdf.

La tabla II muestra el nuevo formato dado por la SEP en el marco del *segundo taller de actualización sobre los programas de estudio 2006* efectuado en varias sedes del Distrito Federal en el mes octubre de 2007. Con este formato se pretende que el docente elabore la planeación del curso de ciencias[7].

**TABLA II.** Campos principales del Formato para Planeación.

| **Subsecretaría de Educación Básica** Dirección General de Desarrollo Curricular **FORMATO PARA PLANEACIÓN** | | | |
|---|---|---|---|
| DATOS GENERALES | | | |
| ESCUELA: | ASIGNATURA: | GRADO Y GRUPO: | NOMBRE DEL PROFESOR(A) |
| BLOQUE: | | PERÍODO | |
| TEMA: | | SUBTEMA(S): | |
| PROPÓSITO DEL BLOQUE | | | |
| APRENDIZAJES ESPERADOS | | | |
| Momentos de organización de actividades | Recursos didácticos | | Orientaciones para la evaluación |
| Actividades de inicio (tiempo _____) | | | |
| Actividades de desarrollo (tiempo _____) | | | |
| Actividades de cierre (tiempo _____) | | | |

Al hacer la comparación de los campos a llenar en el formato de la tabla I correspondientes al plan 1993, con los de la tabla II (RES 2006), se puede observar un incremento de la actividad de planeación.

**TABLA III.** Formato de control para la materia de español de segundo grado de educación secundaria.

| TIEMPO | ACTIVIDADES Y TEMAS DE REFLEXIÓN | RECURSOS | OBSERVA-CIONES |
|---|---|---|---|
| Aspectos a Evaluar | | Indicadores | |
| Proceso | | | |
| Producto | | | |

Anteriormente, al hacer la planeación, las preguntas base eran:
- ¿Qué quiero lograr? Para la columna de propósitos.
- ¿Cómo lo voy a lograr? Para la columna de estrategias.

- ¿Con qué lo voy a lograr? Para la columna de recursos didácticos y materiales.
- ¿Cómo lo voy a evaluar? Para la columna de procedimiento e instrumento de evaluación.

Ahora la planeación en la RES, sin más explicaciones, incluye además de los espacios para anotar los propósitos, los recursos didácticos y los instrumentos de evaluación, un espacio para considerar las actividades de inicio, de desarrollo y de cierre para cada subtema.

Se ha hecho especial énfasis por parte de las autoridades en centrar la planeación en los aprendizajes esperados. Estos son el eje principal de toda actividad de enseñanza aprendizaje dentro o fuera del aula.

La RES persigue un trabajo por proyectos interdisciplinario y el desarrollo de competencias, por lo que los aprendizajes esperados son la columna vertebral de la planeación dado que son los indicadores de desempeño que sustentan a las competencias[8].

Los aprendizajes esperados para cada asignatura son dictaminados por la SEP. El docente podrá decidir sobre las estrategias y recursos a implementar, el cómo y el con qué.

Resulta trascendente destacar que muchos docentes no sólo están aprendiendo a implementar una planeación más completa, sino una nueva forma de trabajo que, para algunos, representa un cambio radical en su tradicional práctica docente.

Cabe señalar que en cada materia se han distribuido formatos diferentes acordes a las necesidades y requisiciones de cada inspector o jefe de clase. Estos formatos pueden encontrarse y compararse en la página de Internet del Instituto de Humanidades y Ciencias de la Cuidad de México, en la sección de maestros[9].

Esta situación generó cierto descontrol y malestar entre docentes de diferentes materias pues al intentar consultarse unos a otros sobre el llenado de formatos no encontraron un patrón universal como anteriormente se tenía. Para ilustrar esta situación podemos citar el caso de la materia de Español para segundo grado cuyo formato contiene entre otras, las columnas que se muestran en la tabla III.

Y el caso de la materia de Historia de segundo grado cuyo formato se muestra en la tabla IV.

**TABLA IV.** Formato de control para la materia de historia de segundo grado de educación secundaria.

| Aprendizaje(s) esperados(s) (según el programa) | | |
|---|---|---|
| Conocimiento histórico | Habilidades (procedimientos) | Actitudes y valores |
| | | |

---

[7] SEP (2007). Segundo taller de actualización sobre los programas de estudio 2006. Ciencias II. Guía de Trabajo. Anexo 6, secuencia didáctica p. 91.

[8] Cita de la referencia [5]: *"Una vez definida la competencia se definen los indicadores de desempeño. Un indicador de desempeño es un descriptor del proceso para llegar a adquirir la competencia... En secundaria se llaman aprendizajes esperados"*

[9] Instituto de Humanidades y Ciencias. Sitio web http://www.inhumyc.edu.mx/profes.html, visitada el 18 de enero de 2008 a las 14:15.

*Alfonso Cuervo, César Mora y Ricardo García-Salcedo*

| Aspectos a evaluar | |
|---|---|

Además de la planeación antes mencionada, la SEP ha solicitado que los docentes de Ciencias, elaboren secuencias didácticas bajo el formato mostrado en la figura 1a y figura 1b[10].

La diversidad de formatos empleados para la planeación, de acuerdo a los requerimientos y necesidades de cada materia, hace ver que el diálogo entre todos los autores de la reforma es discordante por momentos y que estamos todavía en una etapa de implementación susceptible de ajustes.

Todos estos formatos, lejos de haber sido diseñados desde estrategias probadas y documentadas o de haber sido el resultado de una implementación probada previamente, solo se han hecho llegar hasta las manos de los docentes para que los implementen sin siquiera algún instrumento para su evaluación. Cabe señalar que los docentes adaptados o no a los nuevos formatos de trabajo tampoco dan muestras de requerir comprobación alguna sobre la eficacia de los nuevos formatos.

El resultado de esta conformación multidisciplinaria de distintos actores para crear e implementar la RES puede verse reflejado en el sitio de Internet de la reforma http://www.reformasecundaria.sep.gob.mx/ que muestra, para la mayoría de las materias, un sitio web con orientaciones y recursos didácticos variados, como el estudio de caso para Geografía, las sugerencias didácticas para Español y las secuencias didácticas para Ciencias, que por cierto no siguen el formato distribuido en las juntas de academia.

Es notable que en materia de planeación, la Reforma implique un quehacer programado y estructurado al que algunos docentes consideran como trabajo burocrático y otros ven como un gran reto por la tradicional forma de llevar a cabo su práctica docente.

De esta multidisciplinaria forma de trabajar en cada materia se pretende lograr una interdisciplinaria forma de actuar en los alumnos y nos hemos detenido a tratar el tema de los formatos de planeación de clase porque son la base con la que se desarrolla el trabajo dentro y fuera del aula y porque son un elemento indispensable para dar el seguimiento por parte de las autoridades de la SEP a la evolución y adaptación de la RES en las escuelas.

Otras medidas de seguimiento están detalladas en los informes nacionales. El cuarto informe nacional correspondiente al ciclo 2006-2007, elaborado por la SEP [6], arroja resultados muy positivos en lo que a implementación de la Reforma Educativa se refiere, pero cabe destacar que el mismo informe cita que el seguimiento a las escuelas en el ciclo escolar 2006-2007 incluye dos vertientes:

---
[10] SEP (2007). Segundo taller de actualización sobre los programas de estudio 2006. Ciencias II. Guía de Trabajo. Formato disponible en sitio web del Instituto de Humanidades y Ciencias http://www.inhumyc.edu.mx/maestros/FORMATOS%20DE%20PLANEACION%20DIDACTICA/ciencias/formato%20plan%20ciencias%20II.doc.

1) Se continúa haciendo seguimiento a las 127 escuelas participantes en la primera etapa de implementación
2) Se inicia el seguimiento a por lo menos cinco escuelas secundarias más por entidad de las tres modalidades –general, técnica y telesecundaria–, en el marco de la generalización de la reforma [6].

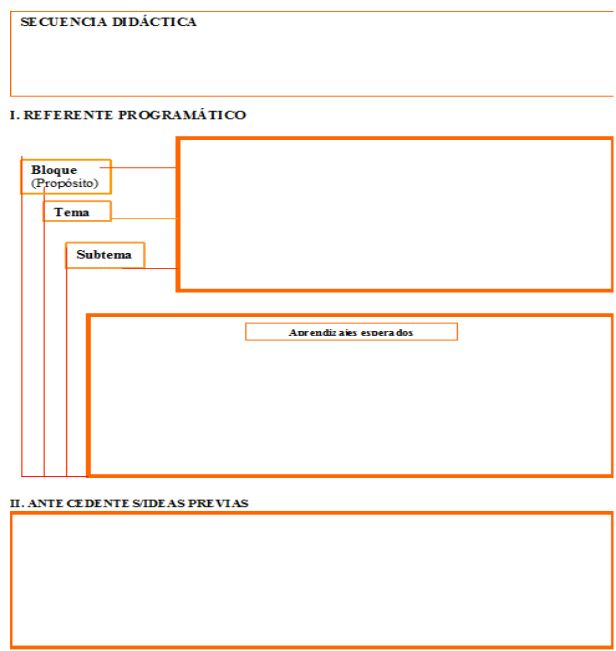

**FIGURA 1a.** Formato de secuencia didáctica solicitado por la SEP.

Estas cifras dejan claro que el cuarto informe nacional toma una muestra representativa amplia pero no abarca el total de las escuelas del país.

El siguiente párrafo es un extracto del citado cuarto informe nacional [6]:

*"Los maestros declaran tener una preocupación que atañe a su persona respecto a la Reforma: la carencia de capacitación para el manejo de los enfoques y contenidos de los programas de estudio, así como para atender la planeación del trabajo por proyectos; la posible resistencia al cambio de algunos docentes, o bien la falta de compromiso para cambiar su práctica docente y los retos que le representa la Reforma misma".*

El mismo informe reconoce que cerca de un 30% de los directores de escuela consideró poco adecuada o inadecuada la información tratada en Taller de Inducción sobre la Reforma Educativa, donde se analizaron y discutieron aspectos de organización general de la escuela, del fortalecimiento de la práctica docente y elementos de apoyo para la resolución de dudas de los maestros.

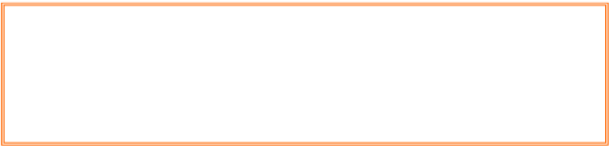

**FIGURA 1b.** Formato de secuencia didáctica solicitado por la SEP.

Contrasta con la visión hasta aquí expuesta, el que en el tercer reporte titulado *"Comentarios de los docentes sobre los programas de estudio de 2º grado"*, así como en el reporte titulado *"Percepciones y valoraciones de los docentes y directivos sobre el Plan de Estudios 2006 con sus respectivos programas y el Modelo Renovado de Telesecundarias"*; ambos pertenecientes al quinto informe nacional sobre seguimiento a las escuelas en el ciclo 2006 – 2007[11] elaborado por la SEP, se vislumbra un panorama de adaptación y convencimiento por parte de los docentes, destacando que sus preocupaciones oscilan más hacia el tema de la infraestructura y los recursos materiales. 35% de los docentes que han brindado información para la elaboración de este quinto informe señalan que para mejorar la aplicación de los programas de estudio, se requiere de capacitación en materia de planeación, diseño y diversificación de situaciones y actividades didácticas.

Es evidente entonces, que tanto el tiempo como los materiales abordados en el segundo taller de actualización sobre programas de estudio 2006, han sido insuficientes. En el citado taller de actualización se ha analizado la guía de trabajo[12] que propone en sus anexos cinco ejemplos de secuencias didácticas para abordar un tema correspondiente a cada bloque.

Para los docentes que asistieron a éste taller y atendieron la propuesta de trabajo a través de las secuencias sugeridas resulta un reto considerable alcanzar tanto en tan breve tiempo, y por esta razón en el quinto informe se da el resultado anteriormente citado en relación a la necesidad que expresan los docentes de contar con capacitación en planeación, diseño y diversificación de situaciones y actividades didácticas.

La siguiente tabla (tabla V), muestra un ejemplo que resume la propuesta de trabajo abordada por la guía de trabajo de la SEP para el caso de los bloques 1 y 2 de la materia de Ciencias II, la cual evidencia la cantidad de aprendizajes esperados y las horas de clase en que éstos deben ser alcanzados. En los ejemplos de secuencia para los bloques 3 y 4, la guía de trabajo pretende alcanzar 7 y 6 aprendizajes esperados respectivamente en un tiempo de 6 horas. Para el bloque 5, la guía de trabajo no indica un tiempo total.

**TABLA V.** Relación de aprendizajes esperados en horas de clase.

| Secuencia Didáctica para: | Aprendizajes esperados en esta secuencia | Horas totales |
|---|---|---|
| Bloque 1: El Movimiento. Tema: El trabajo de Galileo: Una aportación importante para la ciencia Subtema: 2.2 ¿Cómo es el movimiento cuando la velocidad cambia?<br>• La aceleración<br>• Experiencias alrededor de movimientos en los que la velocidad cambia.<br>• Aceleración como razón de cambio de la velocidad en el tiempo<br>• Aceleración en gráficas velocidad-tiempo | ✓ Identifica a través de experimentos y de gráficas las características del movimiento acelerado.<br>✓ Aplica las formas de descripción y representación del movimiento analizadas anteriormente para describir el movimiento acelerado.<br>✓ Identifica la proporcionalidad en la relación velocidad-tiempo.<br>✓ Establece la diferencia entre velocidad y aceleración.<br>✓ Interpreta las diferencias en la información que proporcionan las gráficas de velocidad-tiempo y las de aceleración-tiempo provenientes de la experimentación o del uso de recursos informáticos y tecnológicos. | 6 horas |
| Bloque 2.- Las Fuerzas Tema: 2. Una explicación del cambio: La idea de fuerza. Subtema: 2.1 La idea de fuerza: el resultado de las interacciones.<br>• El concepto de fuerza como descriptor de las interacciones.<br>• La dirección de la fuerza y la | ✓ Relaciona el cambio en el estado de movimiento de un objeto con la fuerza que actúa sobre él.<br>✓ Infiere la dirección del movimiento con base en la dirección de la fuerza e identifica que en algunos casos no tienen el mismo sentido.<br>✓ Reconoce que la fuerza es una idea que describe la interacción entre objetos, pero | 5 horas |

---

[11] Seguimiento a las escuelas. Ciclo escolar 2006-2007. Quinto Informe Nacional. Secretaría de Educación Pública. Documento disponible online en www.reformasecundaria.sep.gob.mx Consultado el 23 de enero de 2008 a las 02:00 p. 66 y p. 121.

[12] Ciencias II. Guía de Trabajo. Segundo Taller de Actualización sobre los Programas de estudio 2006 Anexos 6 a 10 pp. 91-176.

*Alfonso Cuervo, César Mora y Ricardo García-Salcedo*

| Secuencia Didáctica para: | Aprendizajes esperados en esta secuencia | Horas totales |
|---|---|---|
| dirección del movimiento.<br>• Suma de fuerzas<br>• Reposo. | no es una propiedad de los mismos.<br>✓ Analiza y explica situaciones cotidianas utilizando correctamente la noción de fuerza.<br>✓ Utiliza métodos gráficos para la obtención de la fuerza resultante que actúa sobre un objeto.<br>✓ Identifica que el movimiento o reposo de un objeto es el efecto de la suma (resta) de todas las fuerzas que actúan sobre él.<br>✓ Obtiene la fuerza resultante que actúa sobre un cuerpo y describe el movimiento asociado a dicha fuerza.<br>✓ Relaciona el estado de reposo de un objeto con el equilibrio de fuerzas actuantes sobre él y lo representa en diagramas. | |

## IV. IMPLICACIONES DEL NUEVO PLAN DE ESTUDIOS EN LA MATERIA DE CIANCIAS II

La Física, que se abordaba a lo largo de los tres años de Secundaria en el antiguo Plan de Estudios de 1993, comenzado con Introducción a la Física y a la Química para continuar con Física I en segundo grado y Física II en tercer grado; en distribuciones de 3 horas semanales, se ha concentrado en el segundo grado únicamente llevando el nombre de Ciencias II. La carga horaria es de 6 horas a la semana, distribuidas a razón de 5 horas de trabajo en aula (teoría), por 1 de laboratorio.

**TABLA VI.** Campos y temáticas para el nuevo plan de estudios de Ciencias II.

| Campos de la física | Elementos para la representación de fenómenos físicos | Temáticas |
|---|---|---|
| Estudio del movimiento. | Esquemas descriptivos. | Bloque I. El movimiento. La descripción de los cambios en la naturaleza. |
| Análisis de las fuerzas y los cambios. | Relaciones y sentido de mecanismo. | Bloque II. Las fuerzas. La explicación de los cambios. |
| Modelo de partículas. | Imágenes y modelos abstractos. | Bloque III. Las interacciones de la materia. Un modelo para describir lo que no percibimos. |
| Constitución atómica. | Imágenes y modelos abstractos. | Bloque IV. Manifestaciones de la estructura interna de la materia. |
| Universo interacción de la física, la tecnología y la sociedad | Interpretaciones integradas y relaciones con el entorno. | Bloque V. Conocimiento, sociedad y tecnología. |

El curso se divide en cinco bloques, cada bloque tiene definidos los aprendizajes esperados que deberán ser obtenidos como resultados. Al final de cada bloque se sugieren una serie de proyectos que pongan en práctica lo aprendido y que desarrollen las habilidades (competencias) del alumno. La tabla VI muestra las temáticas de estos cinco bloques [1].

De acuerdo a lo mencionado por jefes de clase e inspectores en los cursos y talleres brindados dentro de las juntas de academia, en el nuevo plan de estudios se pretende dar un nuevo enfoque que pretende ir del estudio del fenómeno a la ecuación que lo describe matemáticamente de tal forma que fomenta el trabajo por proyectos a partir de la elaboración de secuencias didácticas que contemplen los aprendizajes esperados. El siguiente esquema [1] (Figura 2) muestra un cuadro comparativo entre lo estudiado en el plan de estudios del 1993 contra lo que ahora conforma el nuevo plan de estudios.

Puede observarse en la figura 2, que el primer encuentro con la Física anteriormente era a través del tema de las propiedades físicas y su medición.

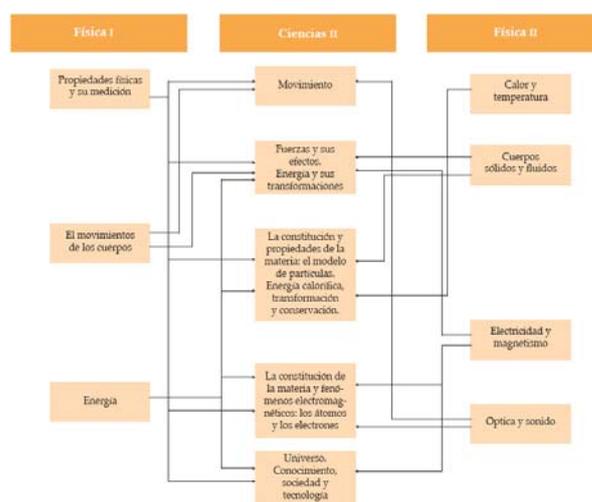

**FIGURA 2.** Relación entre los contenidos de Física I y Física II del plan de estudios 1993 y el nuevo plan de Ciencias II 2006.

Se hacía en aquel entonces la conciencia de la importancia de medir, se introducía al alumno en el manejo de los patrones de medida, el concepto de magnitud, la diferencia entre una magnitud escalar y otra vectorial, el Sistema Internacional de Unidades, etc. Hoy los alumnos entran a la Física por medio del tema del movimiento y uno de los aprendizajes esperados primordiales es que el alumno sea capaz de describir los movimientos en base a la información sensorial. Sin embargo, ahora se tienen que integrar estos conocimientos en todos los bloques, revisando la forma de medir y las unidades útiles dependiendo de lo que se esté estudiando

Al intervenir los sentidos se justifica porque la luz y el sonido que eran abordados hasta el tercer grado de secundaria, son ahora involucrados desde el inicio del curso, ver por ejemplo parte del nuevo programa de estudios que se muestra en la Tabla VII.

El cambio es considerable, algunos docentes apuntan a que la reforma educativa ha *desmatematizado* el programa, le ha restado contenidos indispensables.

Quienes defienden incondicionalmente a la reforma apuntan a que ahora se debe ir del fenómeno a la fórmula y que los temas seguirán siendo cubiertos en su totalidad.

**TABLA VII**. Parte del programa de Ciencias II.

| 1. La percepción del movimiento | **Aprendizajes esperados:**<br>✓ Reconoce y compara distintos tipos de movimiento en el entorno en términos de sus características perceptibles.<br>✓ Relaciona el sonido con una fuente vibratoria y la luz con una luminosa.<br>✓ Describe movimientos rápidos y lentos a partir de la información que percibe con los sentidos y valora sus limitaciones.<br>✓ Propone formas de descripción de movimientos rápidos o lentos a partir de lo que percibe. |
|---|---|

El éxito de este nuevo plan de estudios dependerá de un cambio de actitud por parte del docente, este cambio debe considerar disposición y compromiso. Abolir la falta de preparación de los docentes, especialmente de aquellos que han llegado al aula por buscar una alternativa laboral que complemente su actividad profesional más que por hacer de la docencia una forma de vida.

Dependerá también de que las planeaciones de secuencias didácticas sean congruentes, del uso desde luego de las TIC (Tecnologías de la Información y la Comunicación, por sus siglas) pero sobre todo de estar abierto a trabajar en colaboración con el resto del plantel en un trabajo interdisciplinario.

## IV. CONCLUSIONES

Si bien es cierto que la reforma educativa ha sido más impuesta que propuesta para los docentes en general, es también cierto que los docentes están abiertos a trabajar en ella. Acuden a las juntas y escuchan el llamado que sus jefes de enseñanza e inspectores hacen para trabajar en los nuevos planes de estudio.

Los docentes están abiertos al cambio y ávidos de encontrar una forma que les permita lograr los aprendizajes esperados.

Aprovechar estos tiempos de cambio y de adaptación, es una oportunidad para sembrar en los docentes estrategias de trabajo que mediante secuencias didácticas diseñadas logren los objetivos que la reforma pretende.

El aprendizaje activo de la física tiene hoy una gran oportunidad de desarrollo en México si logramos captar la atención de los docentes y mejorar sus estrategias de trabajo.

Por lo que un trabajo que ya se está realizando como continuación de esta crítica, es diseñar secuencias didácticas a través de diversas estrategias, principalmente el aprendizaje activo de la Física, por ejemplo [8].

## AGRADECIMIENTOS

## V. REFERENCIAS

[1] SEP. Plan de Estudios 2006. Educación Básica. Secundaria. México: SEP. Disponible en formato PDF en http://www.reformasecundaria.sep.gob.mx/ciencia_tecnologia/doctos/programa.pdf (2006). Consultado el 24 de enero de 2008 02:47.

[2] SEByN 2002, Documento Base. Reforma Integral de la Educación Secundaria. 2002. Documento en formato PDF, disponible online en:
http://www.reformasecundaria.sep.gob.mx/doc/docbase.pdf. Consultado el 30 de enero de 2008.

[3] Candela, A. *Comentarios a los programas de ciencias I, II y II en el marco de la RES,* Revista Mexicana de Investigación Educativa **11**, 1451-1462 (2006).

[4] SEP (2006), La orientación y la tutoría en la escuela secundaria. Lineamientos para la formación y la atención de los adolescentes. Disponible en *Asignaturas currículo en línea* http://www.reformasecundaria.sep.gob.mx/. Consultado el 24 de enero de 2008 02:47

[5] Frade, L., *Desarrollo de competencias en educación básica: Desde preescolar hasta secundaria*, (Editorial: Calidad educativa consultores, México, 2007).

[6] Seguimiento a las escuelas. Ciclo escolar 2006-2007. Cuarto Informe Nacional. Secretaría de Educación Pública. Documento disponible online en:
www.reformasecundaria.sep.gob.mx. Consultado el 18 de enero de 2008.

[7] Página: http://www.inhumyc.edu.mx/profes.html. Consultado el 18 de enero de 2008.

[8] García-Salcedo, R. y Sánchez D., *La enseñanza de conceptos físicos en secundaria: diseño de secuencias didácticas que incorporan diversos tipos de actividades*, Lat. Am. J. Phys. Educ. **3**, (2008).